\documentstyle[epsfig]{aipproc}
\begin{document}
\title{Summary: ZeV Air Showers \\
               Where do we Stand?} 

\author{Enrique Zas}
\address{Departamento de F\'\i sica de Part\'\i culas,\\
Universidade de Santiago de Compostela, E-15706 Santiago, Spain.\\
zas@fpaxp1.usc.es}

\maketitle

\begin{abstract}
Here an attempt is made at summarizing the presentations, most of 
which were about the highest energy particles observed in nature. 
Particular attention is paid to the solutions to the Ultra High Energy 
Cosmic Ray particles, to the new and forthcoming data, to the new 
proposals for experiments and to the role of primary composition, that 
were amongst the most discussed subtopics.   
\end{abstract}

\section*{Introduction}
%
%
%
%

The discovery of events with energies above $10^{20}$~eV dates back 
to the 1960's, to the early days of air shower detection experiments 
\cite{EeVents}. In fact the cosmic ray spectrum has been observed as a 
continuum at all energies since the beginning of the 20th century, 
apparently only limited by the acceptance 
of the existing detectors. The prediction of the 
Greisen-Zatsepin-Kuz'min (GZK) cutoff dates back to the 
same decade of the 1960's \cite{GZK}. The cutoff should appear 
at energies just above $4~10^{19}$~eV because of proton 
attenuation in the Cosmic Microwave Background (CMB) . 
Heavy nuclei are also attenuated in both the infrared and cosmic 
microwave background radiation fields, at roughly the same energies through 
photo-disintegration and pair production. 
The mean free path for these processes is of order 
a few Mpc and even allowance of successive interactions with 
small fractional energy loss rises the attenuation length to 
roughly 50~Mpc. 
The attenuation distances for photons 
in the same background fields are even smaller. 

If these events are coming from 
extragalactic distances, as suggested by the close to isotropic 
distribution in the relatively well known galactic fields, they 
should show the GZK cutoff just below the $10^{20}$~eV range. 
If on the contrary they are not extragalactic, we are facing with an 
unknown source of the highest energies particles 
ever discovered, which is quite close on cosmological scales, 
either challenging dimensional analysis of acceleration processes or 
opening up the way to new physics. A large number of hypotheses have 
been put forward including new astrophysical objects, new particles, 
new interactions or the violation of well established principles. 
The nature of most of these hypotheses, which span a number of research fields 
which are traditionally very far apart, is clearly a sign that we are 
debating a remarkable problem. Nonetheless this problem has resisted 
the efforts of theorists and experimentalists chasing for an 
acceptable solution for over 40 years.  

In this very successful conference we have heard about recent 
developments in the 
field both from the theoretical and experimental sides. 
One of the main ideas that sprouts from it is that a new 
generation of large aperture experiments has just started. It 
will drastically speed up the remarkably low pace in building up 
statistics dictated by fluxes below one particle per square kilometer 
per century. These detectors will also help to determine the primary 
composition. 

In this article I attempt to summarize the material presented in the 
conference. This is by no means an easy task and inevitably I will 
present the field in a subjective way and I will make 
omissions for which I apologize in advance. 
I first briefly comment on the mystery of the ultra high energy cosmic 
rays, to then refer to some of the alternative solutions which were 
discussed that I have divided  
in three groups, those models that require acceleration, those 
models that require fragmentation and decay of massive particles 
and those that avoid the GZK effect. I then discuss the importance of 
composition, one of the issues that was most addressed in the conference. 
Finally I refer to the experiments that were discussed and conclude. 

\section*{The Post GZK Flux Puzzle}

The high energy tail of the measured cosmic ray flux, 
those particles arriving well above the energy of the 
Greisen-Zatsepin-Kuz'min cutoff, 
presents a complex challenge that is 
still unresolved as was pointed out in many of the talks 
\cite{biermann,blasi,cronin,ellis,gaisser,harari,selvon,linsley,masperi,olinto,sigl,swain,weiler}.  
If we conservatively assume that the high energy end of the 
cosmic ray spectrum is due to well established particles 
such as protons or nuclei (that constitute the low energy end of the 
cosmic ray spectrum) or even photons, we can be fairly confident that 
their interactions with magnetic fields and background radiation 
fields are well understood. If these particles are coming from 
distances exceeding a few tens of Mpc the observed flux should 
have an imprint of the interactions with the background radiation 
fields, that is it should display the GZK cutoff at about $4~10^{19}$~eV.  

The actual measurements are quite limited to the primary energy spectrum, 
the arrival directions and some information related to the nature of 
the arriving particles (mass composition) based on shower development. 
As more data has accumulated over the years from different experiments 
and efforts have been made to analyze the combined data \cite{alan} 
it has become rather clear that: 
a) The data exhibit a hardening of the spectral index at an energy 
of $8~10^{18}$~eV \footnote{There is however data from the Fly's Eye 
experiment, using the fluorescence technique, that suggests the change of 
slope occurs at about half this energy.} and 
b) That there is no evidence of the GZK cutoff, with data reaching 
$3~10^{20}~$eV. \footnote{The highest energy event seen with 
the fluorescence technique provides a calorimetric measurement 
of shower energy, reinforcing the results obtained 
with particle detector arrays.} 
The change of spectral index is very 
suggestive of a different component of the spectrum above 
$10^{19}$~eV, the Ultra High Energy Cosmic Rays (UHECR) which provide 
the central topic of this conference. 
Throughout this article I will also refer to such a component 
as the post GZK particle flux. 

According to the observations these particles 
are unlikely to be 
coming from distances exceeding about 50~Mpc. This 
distance scale was discussed as an observational "horizon" for protons 
(or nuclei) because of their interactions with the background 
fields \cite{linsley}.  
So far only two mechanisms have been suggested by which 
particles can attain the highest energies, namely: 
a) Direct acceleration of charged particles 
and b) Fragmentation of the decay products of other particles. 

If the particles are accelerated, 
the exceptional energies achieved are very 
demanding on possible sources that could accelerate them. The 
constraint comes from basic dimensional arguments. The acceleration 
of a particle of charge $Ze$ to an energy $E$ requires a minimum value 
for the product of the size of the accelerator $L$ 
and its magnetic field $B$, namely:
\begin{equation}
 E < ZeBcL
\label{Ebound}
\end{equation}
where $c$ is the speed of light. 
This is traditionally illustrated by the Hillas plot \cite{hillas} 
and several versions of it have been discussed at the conference 
\cite{cronin,olinto,sigl,selvon}. 
Few of the astrophysical objects and structures known satisfy 
the constraint. Moreover those that do must be very efficient in 
reaching the maximum energy and many of them are either too large 
or too distant compared to the 50~Mpc horizon. There are however 
possible acceleration sites that cannot be excluded 
yet. 

The difficulties in acceleration scenarios has opened the 
way to alternative proposals in which the post GZK particles 
either avoid the interactions with the CMB or they are postulated to 
be a secondary flux produced locally from the decay of 
other particles. 

On pretty general grounds all models for the origin of the 
post GZK particle flux are very dependent on at 
least three assumptions in a very interrelated fashion, namely:
\begin{itemize}
\item Hypothesis about the 
possible production sites for these particles which determine both the 
source distributions and the distance travelled before reaching us. 
\item Assumptions about the nature of the particles 
themselves which determine their interactions with the magnetic fields 
and background radiation fields in their path to us. 
\item Models for magnetic field distributions in the Galaxy, galactic halo, 
clusters of galaxies and intergalactic space that condition the 
distribution of arrival directions.
\end{itemize} 

Part of the puzzle lies in the 
fact that it is very difficult to extract unconditional conclusions with 
the limited available data. Each model has to be tested against observations. 
The interpretation of the information on the post GZK particle 
arrival directions is completely dependent on these three hypotheses. 
Moreover the knowledge of the magnetic fields are pretty limited 
outside the light-matter distribution of our galaxy. 
As an example it was pointed out that it is possible that the 
magnetic field distribution provides strong flux magnification and 
depletions in preferred directions, that could explain 
the apparent absence of the GZK cutoff and the observed multiplicities 
in the arrival directions \cite{harari}. 

In a way the difficulties associated with this intricate 
interdependence of assumptions 
makes these particles so attractive to different fields 
of research including particle physics at extreme 
energies, astrophysical objects and the study of extragalactic 
magnetic fields. 

\section*{Models}

The different types of alternatives for the origin of the UHECR 
were reviewed 
by Angela Olinto \cite{olinto} who classified them in two main groups: 
Those that push these conventional acceleration ideas to 
extremes in order to accommodate the data or {\sl Zevatrons} 
and those that invoke new physics. 
Particular attention was paid in her review 
to the differences in the arrival direction distributions, the 
spectral features and primary composition that can be expected from 
different models. It was stressed that most models 
imply strong requirements on the magnetic fields to fit the 
observations \cite{olinto}. 
A number of proposals were critically discussed by 
several speakers. I will select a few.

\subsection*{Standard Fermi Acceleration Models}

Stochastic particle acceleration in the interaction of particles 
with astrophysical shocks is the conventional astrophysical answer 
to the question about the origin of the cosmic rays below the GZK 
cutoff. The mechanism can be extended to the Zevatron models to 
also explain the origin of the post GZK flux. 
These models are also called "bottom-up'' 
scenarios in contrast to a solutions which avoid acceleration by 
assuming particles are created with high energy already, the so called 
"top-down'' models. 

A first general approach to these models was made by Thomas 
Gaisser \cite{gaisser} on a energy 
balance argument. The power density needed to be injected in cosmic rays 
to produce the higher energy region of the cosmic ray spectrum can be 
computed for a given type of object with a known distribution and assuming 
a nominal spectral index of $\gamma=2$ characteristic of Fermi 
acceleration, mimicking the well known argument supporting Fermi 
acceleration in Super Nova Remnants for the origin of the bulk of 
cosmic rays. It turns out that 
for galaxies, clusters of galaxies, 
Active Galactic Nuclei (AGN) and Gamma Ray Bursts (GRB) the UHECR 
power requirement is a reasonably small fraction of the 
power density emitted by each of these classes of objects. 
The argument can 
be reconsidered for strong relativistic shocks such as those expected 
to be found in AGN's and GRB's, which typically result in steeper 
spectral indices $\gamma \simeq 2.2-2.3$. 
In these shocks the particle can be accelerated provided it achieves 
a minimum energy by some other means. 
The resulting power balance is very dependent on both the spectral index 
and the injection energy. Part of the extra power that would be 
required because of the steeper spectrum is compensated by an 
increased injection energy. 
Simple energetic considerations open up these possibilities and do 
not allow much discrimination between them \cite{gaisser}. 


Fermi Acceleration in Gamma Ray Bursts has been suggested as a 
possible UHECR source \cite{waxman}. Although the energetics 
may be adequate, the implied assumption that the bursts have a 
comoving density which is independent of redshift is somewhat 
contrived \cite{stecker}. We should probably have to wait for a 
better understanding of these interesting phenomena before this 
possibility 
can be critically revised. In this respect we heard of an interesting  
proposal for GRB observations which could shed some light into the 
general GRB problem \cite{oscar}. 

Powerful radiogalaxies are certain candidates for post GZK acceleration. 
It was stressed that these objects provide the largest shock waves 
known and that the standard radioastronomical observations already 
demand highly energetic particles to explain the energy transport 
along the jets \cite{biermann}. It has been claimed 
that a single source could provide the solution of the UHECR problem 
\cite{ahn}. 
It is possible to devise reasonable magnetic field models 
in which backtracking of post GZK particles under the assumption of 
mostly proton primaries and one helium nucleus, leads to M87, 
a nearby radiogalaxy in the Virgo cluster \cite{ahn,biermann}. 
These hypotheses will be further tested once the statistics 
builds up. 
Very interesting results for magnetic field flux magnifications 
and spectral deformations in general and for this particular 
model were presented \cite{harari}. 
These local magnifications could significantly affect power 
requirement estimates and the arguments based on them. 

The shape of the spectrum at the cutoff is a signature of the  
source distribution and an indirect handle on the distance to the 
production sites. If the post GZK particles are of extragalactic 
origin but they are produced at higher rates in a region close to us 
relative to the 50~Mpc proton horizon, the GZK cutoff effect is 
mitigated \cite{berezinskii,bahcall}. 
A recent simulation of this effect was discussed at the conference 
\cite{blasi} assuming the post GZK flux is correlated with 
the dark matter distribution. The spectral index assumed for 
the injection spectrum becomes important and the observed flux 
above $10^{20}~$eV would require a hard spectral index for injection 
with $\gamma \sim 2$. 
A distinct signature for this scenario is the anisotropy of arrival 
directions. Certainly if local density enhancements are responsible 
for the post GZK particles the arrival directions should map these 
density enhancements. 
Large magnetic fields would be needed to explain observed distribution 
of arrival directions but this is consistent with what we 
know about magnetic fields \cite{biermann}. 
Some claims for an excess in the direction of the supergalactic plane 
have been made \cite{stanev}, but the level of confidence is still low 
because the statistics is poor. 

A recent model for acceleration of iron like nuclei in the galaxy was 
addressed \cite{olintons}. These nuclei are stripped off the neutron 
star surface with high magnetic fields that allow acceleration 
to the observed high energies. As the magnetic flux of a 
rotating neutron star reaches the light cylinder it can be converted 
into kinetic energy of the particles in a relativistic wind. If the 
magnetic field is high and the neutron star has a rapid rotation 
the acceleration can reach GZK energies. This alternative 
demands large galactic 
magnetic fields to reproduce the small enhancement in the direction 
of the galactic plane. The enhancement should increase as the energy of 
the observed particles rises. A particularly distinctive characteristic 
is the spectral index at production which is $\gamma \simeq 1$ \cite{olinto}. 
Such a hard spectral index puts most of the energy in the high energy end 
of the spectrum and new experiments extending the energies to the ZeV 
region should be able to measure the spectral index. 


\subsection*{Alternative Solutions}

Most of the solutions avoiding Fermi acceleration in nearby objects 
are motivated by physics beyond the standard model and would 
correspond to new physics in Angela Olintos's review \cite{olinto}. 
As regards the observations from the expected fluxes two very different 
categories emerge. In one class of alternatives the post GZK flux  
involves standard model particles from the decay and fragmentation 
of other particles. A second class avoids the cutoff mechanism either 
with new particles or with standard model particles that have 
unexpected behaviours. 

\subsubsection*{Fragmentation Origin}

Most of the alternative models share the feature of producing the
bulk of the post GZK particle flux by fragmentation into pions.
These models span very different scenarios, for instance the pions
can be due to quarks which are in turn the decay products of a
more massive particle such as a light electroweak
$Z$ boson \cite{weiler}, a Wimpzilla \cite{wimpzilla} which is a 
non thermal long lived particle, or a much heavier $X$ particle 
\cite{hillshramm} from a possible extension of the Standard Model 
into a Grand Unification Theory (GUT). 
These decaying particles are conjectured to be produced in a variety
of mechanisms including, primordial origin, couplings to gravity,
annihilations of topological defects and 
local interactions of distant ultra high energy neutrinos, 
some of which were addressed by several speakers
\cite{olinto,sigl,weiler,ellis,masperi}.
The produced massive particles usually decay into quarks
which fragment into hadrons, mostly pions that in turn decay
into photons and neutrinos.
There are models however in which the quarks can be emitted
directly, for instance in primordial black hole evaporation
\cite{macgibbon}.
As a final result the fragmentation and decay
chain eventually ends up as photons, neutrinos and protons and no
heavy nuclei can be expected.
Such different scenarios ultimately predict roughly similar relative
rates
for the different particles because they rely on the same production
mechanism.

Annihilation of topological defects were historically one of the first 
mechanisms postulated as a source of 
massive particles that give rise to post GZK particles as fragmentation 
products \cite{hillshramm}. Under the heading of topological defect a 
large variety of objects can be included such as monopoles, 
cosmic strings \cite{bhatacharjee}, 
vortons which are classically stable loops of superconducting cosmic 
strings \cite{masperi,masperi2}, and combinations of these 
such as necklaces that combine strings and monopoles \cite{vilenkin}, or 
monopole-antimonopole pairs connected by a string \cite{blanco} 
just to name a few. A number of reviews have dealt with these objects 
\cite{sigl,siglbhat}. 
The possibilities are already heavily constrained by various 
arguments and observations, such as the expected abundances, their 
effects on Big Bang Nucleosynthesis, and the expected secondary fluxes 
of photons and neutrinos \cite{sigl,siglbhat}. 

A particularly important feature of these alternative sources concerns 
their cosmological distributions. 
The possibility that the post GZK particles are locally produced 
in our galactic halo following the fate of cold matter is recently 
gaining more attention \cite{blasi,berezinskycluster,tanco}. If the 
clustering is on scales smaller than the 50 Mpc scale the GZK 
cutoff effect on the spectrum will be mitigated. In many of the 
proposed models involving fragmentation the sources are expected 
to be clustered. An interesting signature of the 
clustering sources is the expected dipole asymmetry associated to 
our position not being central in the halo and clearly related to 
the halo size. 

\paragraph*{Clustered sources:}
One possibility is that the post GZK particles are due to decays of 
heavy metastable particles that cluster in our galactic halo following 
the fate of cold matter and eventually decay. 
They are mostly motivated by and 
different aspects of string and M-theories. 
Some of these were addressed in the conference for 
instance {\sl pentons} motivated by M-Theory compactifications and  
{\sl cryptons} that are supermassive bound states from 
string hidden sectors \cite{ellis}. 

Another model predicting a locally enhanced cosmic ray density is 
motivated by the recent findings about neutrino oscillations, and gives 
qualitatively similar predictions. The model was discussed in several 
talks and involves the production of 
$Z$-bosons in neutrino-antineutrino annihilations as discussed 
by several speakers \cite{weiler,sigl}. 
The model requires a very high energy neutrino flux that could originate 
far beyond the proton horizon of 50~Mpc and a target which is provided 
by massive relic neutrinos that could cluster in our galactic halo. 
No assumptions are made on the origin of the neutrino beam but some 
suggestions have been made \cite{postGZKnu}. 
The maximum neutrino energy has to be pushed even further 
firstly because the observed particles are secondary products of the neutrino interactions and secondly because the neutrinos themselves can also be 
expected to come from decays of other particles. This model is subject to 
important restrictions because of neutrino flux bounds mainly from the 
absence of observations of horizontal or upward going air showers 
\cite{BlancoPRD}. 

All these models have become less likely in the view of very recent studies 
of composition for energies above $10^{19}~$eV, that put the first limits on 
photon abundance \cite{zas,zasbound}. 
The restriction is strongest for models in which the sources are 
clustered in the vicinity of our galaxy so that the photon flux 
does not travel long enough distances to become relatively suppressed 
with respect to the proton flux. 

\subsubsection*{Solutions without Fragmentation} 

A number of solutions to the Ultra High Energy puzzle has been 
suggested invoking particle behaviours that avoid the GZK cutoff. 
Some are familiar standard model particles that develop unexpected 
behaviours at large energies. They include hadronic-like 
cross sections for neutrinos that can mimic proton interactions 
in the atmosphere \cite{bordes,ralston}. Alternatively it has 
also been suggested that Lorentz invariance could be broken 
\cite{mestres,glashow}. 
Other solutions invoke 
new stable particles usually motivated by supersymmetry, 
also called {\sl uhecrons} which have 
higher threshold energies in their interaction with the background photon 
fields but but do interact with matter \cite{farrar,ellis}. 

Particular attention was paid to the recent suggestion that 
Lorentz invariance could be broken. At these extreme 
energies space-time can have a non trivial structure because 
of quantum fluctuations due to the recoil of the vacuum. 
High energy particles see distortions of the metric, a {\sl spacetime 
foam} in which their speed becomes reduced. The Lorentz invariance 
break implied prevents particle production and particularly the 
photoproduction process that is mostly responsible for the 
GZK cutoff \cite{ellis}. Most interesting where 
the first experimental constraints on these ideas searching for 
correlations of delays and distances in the radiation received 
form GRB's in different energy bins. The observations imply that 
the possible quantum-gravity scale $M$ has a lower bound 
$M > 10^{15}~$GeV. Lorentz symmetry violations were also suggested 
from a CP-violating kinematic structure \cite{ahluwalia}.  

Lastly there are solutions involving other particles such as heavy or light 
monopolia \cite{kephart,weiler,swain} or dust grains. 
In the monopole models the monopoles themselves 
would induce the observed air showers, which have however a distribution 
of arrival directions which strongly disfavours such a hypothesis 
\cite{escobar}. The dust grain alternative is ruled out by the 
shower development curve observed for the highest energy event 
\cite{kolb} 
but there is no knowledge of the behaviour of 
monopole induced showers \cite{weiler}. 

\section*{Composition}

Composition was discussed by several speakers as a fundamental 
tool to distinguish between different models which is crucial 
for the correct interpretation of the distribution of arrival 
directions which is mass dependent. 

Particle production in models relying on decays of 
more massive particles is governed by the fragmentation processes.  
If the bulk of the post GZK particles are due to 
fragmented hadrons from heavier particle decays, the relative fluxes 
of different particle species can be 
extrapolated from lower energy fragmentation processes that are 
well known from $e^+ e^-$ collisions in accelerators. 
In fragmentation processes pions are typically produced at a rate 
which is of order ten times higher than protons. Photons arise from 
neutral pion decays while neutrinos are mostly produced in the decays 
of charged pions. The fragmentation and decay chain process is thus 
expected to produce a significantly larger fraction of photons and 
neutrinos than protons (neutrons decay and end up as protons) 
and no heavy nuclei. 
\footnote{Incidentally it was pointed out during this conference 
that these extrapolations are 
not free of uncertainty and that they should be carefully reconsidered 
for instance in connection to supersymmetric extensions of the standard model 
\cite{sigl,blasi,ellis}.} 

One must note that measurements of these fluxes have already proven 
to be very useful in constraining production models 
\cite{sigl,gaisser}. 
These neutral particles if observed will provide non deflected 
information on the fluxes, thus directly reflecting the source 
distribution anisotropy independently on assumptions about the 
magnetic fields which would be a most valuable piece of information. 

As a result one can roughly expect flux ratios of order ten for 
photons and neutrinos with respect to protons. 
The photon ratio is further 
modulated by interactions with the background radiation fields. 
If the model produces more particles near the Galaxy, 
the photon fraction is of order 10 
\cite{ellis,sigl,olinto,selvon} 
while this value can be modified depending on distance to the 
source and fragmentation details to drop to values about 1-3 
when the average 
distance travelled by the particles is in the scale of tens of 
Mpc \cite{kachelriess,sarkar,sigl}. 
On the other hand the neutrino flux will not 
be attenuated and the neutrino to proton flux ratio can be expected 
to be of order ten or higher if there is proton attenuation. 

One must remark that the predictions for the secondary 
fluxes are rather sensitive to proton, electron and photon 
absorption both at the production site and during transport. 
If photons are attenuated, pair production and synchrotron 
losses dump the photon energy density into the MeV region of the 
gamma ray spectrum, linking it to the unknown magnetic 
fields \cite{protstan,sigl}. 

Alternatively stochastic acceleration mechanisms require charged 
particles. If the accelerated particles are heavy nuclei 
higher energies can be achieved. 
The possibility that these ultraenergetic particles are heavy 
nuclei is attractive on a double basis because these nuclei are 
more easily isotropized in a given magnetic field and because 
acceleration models are less constrained by Eq. \ref{Ebound}. 
Nuclei heavier than iron such as gold  
have been also proposed as a plausible solution \cite{swain}. 
Although the mean free path for interactions with the background 
radiation fields is small, the emerging depleted nuclei can undergo 
successive interactions and reach distances in the 100~Mpc range 
keeping a substantial fraction of the original energy \cite{stecker}. 
Heavy nuclei thus relax the constrains on both source distance and 
intensity of the magnetic fields that observations impose. 

If the bulk of the post GZK particles are charged, the study of the 
arrival directions will allow improvements of the current bounds 
on the poorly known magnetic fields outside the galaxy. 
These bounds will be stricter if composition can be determined. 
There are models with heavy nuclei involving both galactic and 
extragalactic sources for which magnetic effects should be most 
important. 
They will be tested against future data on both 
the energy spectrum and anisotropy measurements.  

Acceleration models are also expected to produce secondary fluxes 
of photons and neutrinos. The mechanism is the interaction of the 
protons with matter or radiation and this can happen both at the 
source or during transport. Depending on details of the models 
themselves the secondary fluxes will have however different 
magnitudes \cite{gaisser,sigl}. 
Even if no interactions take place at the source itself 
the GZK cutoff mechanism should produce a minimum flux of secondaries 
in these models. In practically all cases however the relative fluxes 
of neutrinos and photons at energies above the GZK cutoff is expected 
to be less than that of protons, unless there is a high attenuation of 
the protons themselves at origin. 

In models relying on fragmentation, 
photons are expected to present a significant fraction 
of photons in the post GZK flux. 
In this respect the measurements of 
photon composition will be crucial because they provide 
an indirect handle on the clustering process for the post GZK 
sources. Neutrino searches however will extend the observation 
horizon to distances well beyond the 100 Mpc range and will confirm 
clustering scenarios.  
Interesting new ideas were reported for establishing the photon 
composition of the post GZK flux which will be further discussed below. 
These ideas will certainly prove extremely valuable in selecting between 
acceleration and fragmentation mechanisms. 



\section*{EXPERIMENT: Exploring the Post GZK Flux}

\begin{figure}[b!] 
\centerline{\epsfig{file=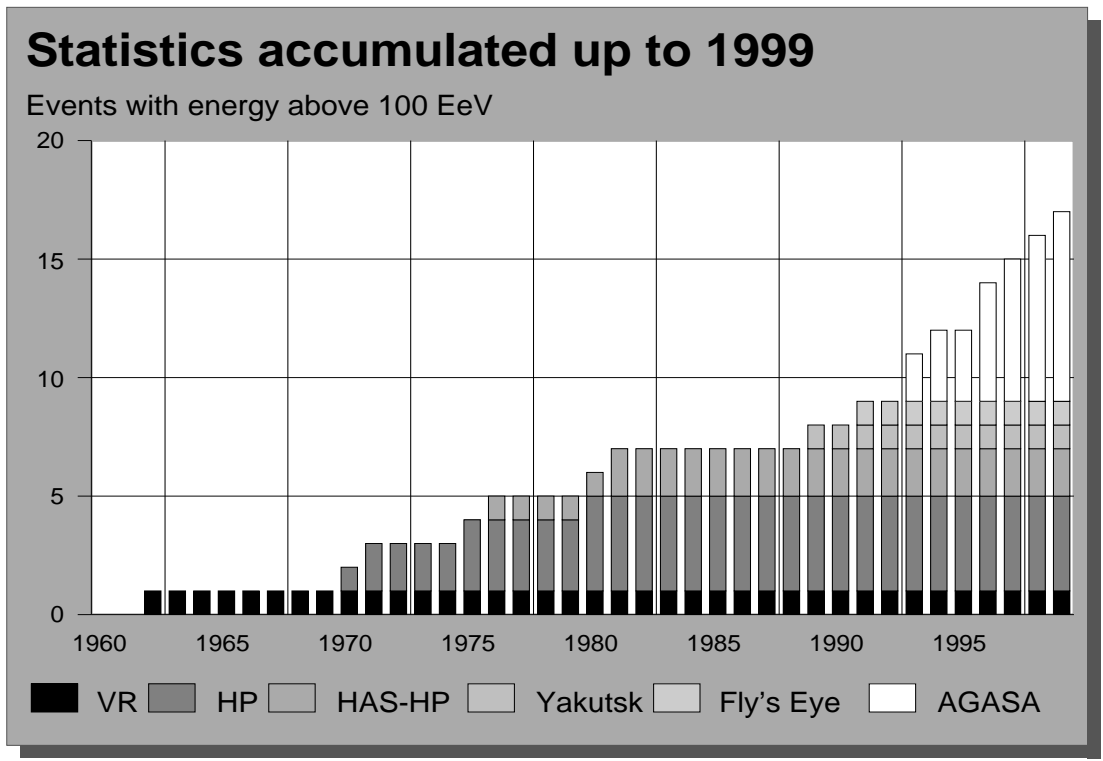,height=4.in,width=5.8in}}
\vspace{10pt}
\caption{Accumulated events of energy exceeding $10^{20}$ eV 
plotted as a funtion of time as detected by different experiments: 
Volcano Ranch (VR), Haverah Park (HP), Horizontal Air Showers in 
Haverah Park (HAS-HP), Yakutsk, Fly's Eye and AGASA.}
\label{gzkevents}
\end{figure}

Altogether only seventeen events with energies exceeding $10^{20}~$eV 
have been reported. Sixteen of them are from four array 
experiments, one from Volcano Ranch \cite{volcano}, six from Haverah 
Park \cite{HP}, one from Yakutsk array \cite{yakutsk} and eight from 
AGASA \cite{AGASA}. The other event, which is the highest energy 
event observed, was detected with Fly's Eye \cite{bird}, a different 
detector concept that measures the fluorescence light from nitrogen 
as the shower develops in the atmosphere. 
The accumulation of events as a function of time is shown in 
Fig.~\ref{gzkevents} including an ultrahigh energy event detected in 
1999 by AGASA \cite{lateagasa}, and two events in a new analysis of the 
Haverah Park data for zenith 
angles above $60^\circ$ \cite{zasrate} which have been discussed in 
this conference. 

Although the statistics are now enough to convince the few remaining 
skeptics, the field clearly demands more data. On the one hand more 
and more precise data on the flux measurements themselves is needed 
to build up statistics for the spectral and anisotropy studies. 
This need is the drive for a number of projects that were presented 
in this conference. Some are already in construction but  
many presentations were concerned about experiments in planning. 
The importance of additional information in the search of the 
origin of the post GZK particles has already been stressed. 
Experiments and techniques that are sensitive to primary composition 
and to secondary fluxes of gamma rays and neutrinos played a central 
part in the conference. 

\subsection*{Existing Detectors}

We heard reports on data obtained with three cosmic ray detectors: 
AGASA, Haverah Park and HiRes. 

\paragraph*{AGASA:}
The status report of this ongoing 
experiment \cite{teshima,lateagasa} was centered on the 8 
events detected above $10^{20}$~eV and on the anisotropy of arrival 
directions. The search of coincidences in arrival directions within 
the angular resolution of the experiment for events with energies 
above $4~10^{19}~$eV results in one 3-fold coincidence and four 
2-fold coincidences with an estimated $0.3 \%$ chance probability. 
The recently reported $10 \%$ anisotropy observed in the region of 
the galactic center and anti-center for energies above $10^{18}~$eV 
\cite{hayashida} was also addressed. 
Although this effect suggests a galactic origin, 
it is observed at a threshold energy well below the GZK cutoff and may 
not have much to do with the origin of the UHECR if as expected they 
indeed are a different component. 

\paragraph*{Haverah Park:}
The interest of Horizontal Air Showers (HAS) was shown not to be limited 
to neutrino detection. We also heard a report on the first analysis of 
inclined HAS induced by UHECR \cite{zas}. A recently developed model for 
muon density maps at ground level \cite{zasmodel} allows event 
reconstruction. As a first application, the analysis of the Haverah Park 
data has resulted in 7 (2) new events above $4~10^{19}~$eV ($10^{20}~$eV) 
\cite{zasrate}. 
 
\paragraph*{HiRes:}
HiRes \cite{hires} \cite{sokolsky} 
is the first large aperture detector that has 
recently started operation based on the nitrogen fluorescence technique. 
The detector has two wide aperture {\sl eyes} and when operating in 
the stereo mode it is expected to measure the energy of cosmic rays 
between 0.1 and 
200~EeV with a $20 \%$ resolution, with improved sensitivity to the 
position of shower maximum and to the determination of the arrival 
directions. 
About five more events with energy above $10^{20}$~eV can be expected 
from the preliminary results of the first six months of data \cite{hires}.  
The forthcoming data analysis of HiRes data is eagerly expected 
because there is only one calorimetric measurement of an event with 
energy above $10^{20}$~eV. It will also help to resolve small 
discrepancies between different experiments. 

\subsection*{Detectors in Planning}

Most advanced plans are for ground detectors using well established 
technologies. The Auger observatory will be the 
first one and the Telescope 
Array and IceCube may be the following ones, but new projects involving 
detection from satellites are already quite advanced in planning and 
could start operating in the second half of the 2000's decade. 

\subsubsection*{Ground Experiments:}

\paragraph*{Auger:} The southern Auger observatory is now in 
construction in {\sl Pampa Amarilla}, Mendoza, Argentina and will 
be the next large aperture detector to become operative 
\cite{cronin}. The engineering 
array is expected to be finished in less than a year time. 
Detailed progress on tank monitoring and calibration 
\cite{salazar}, tyvek reflectivity \cite{zepeda}, 
photomultipliers \cite{fernandez} 
and on Fresnel mirrors for the fluorescence \cite{cordero} was reported.  
The analysis of HAS induced by cosmic rays \cite{zas} 
has shown that the commonly estimated acceptance for cosmic rays with 
zenith angles below $45^{\circ}$ can be doubled for the 
highest energy end of the spectrum.  
As it is a hybrid detector designed to combine the fluorescence and 
the particle density sampling techniques, and it is the first one 
that will perform such hybrid measurements for 
showers with energy exceeding $10^{19}~$eV,   
it will be most adequate for cross calibration of the two different 
techniques and will provide key information to the development of 
the field. 

\paragraph*{Telescope Array:} The Telescope Array (T.A.) is a  
project consisting on an array of eight stations each with 42 
3-m diameter mirrors covering the solid angle from the horizon to 
$58^\circ$ of zenith angle and having a similar aperture to the 
Auger observatories. 
It will have $6\%$ energy resolution and angular 
resolution of $0.6^{\circ}$ \cite{teshima}. 
Decisions on its funding are expected within a year from now. 

\paragraph*{IceCube:} 
The IceCube proposal \cite{gaisser,icecube} 
is a well advanced project aiming to  
expand the proven AMANDA \cite{amanda} technology to a full scale detector 
instrumenting one km$^3$ of deep clear ice. The 
expected rates of neutrinos depend both on the production mechanism 
and on the attenuation of cosmic rays during transport. The mechanism 
determines the ratio of neutrinos to protons and photons. Depending 
on the maximum energy of the neutrinos the flux normalization can 
span a wide range of values in plausible scenarios. 
The expected rates in IceCube in many of these scenarios are at the 
detectable rate \cite{gaisser}. 

\subsubsection*{Detectors from Space:}


Observation of the fluorescence light induced by extensive air showers 
from a satellite is a promising alternative. The thin upper atmosphere 
allows distant detection and a large aperture can be achieved by placing 
the satellite at the planned orbit of the International Space Station (ISS) 
of about 400~km. Several initiatives were discussed in a number of 
presentations. These systems are sensitive both to UHECR and to UHE 
neutrinos by looking for deeply penetrating air showers. 
The \v Cerenkov light reflected from the clouds, earth or the sea can 
be used for distance calibration \cite{scarsi,khrenov}. 

\paragraph*{EUSO:} The Extreme Universe Space Observatory (EUSO) 
\cite{scarsi,catalano,cordero} is a multinational joint effort which 
has been approved by the European Space Agency for an accommodation study. 
It is planned to have a field of view of $60^{\circ}$ spanning 
$200 \times 200$~km$^2$, Fresnel lens optics \cite{catalano,cordero} 
and a finely 
segmented focal plane that will allow the registration of the 
fluorescence light from air showers. It could start operation in 2006 
and is scheduled to operate for three years. 

\paragraph*{KLYPVE:} Another version of the same technique 
adapted to the Russian section of 
the ISS is provided by the KLYPVE project \cite{khrenov,panasyuk,garipov}. 
It has a smaller field of view of $15^{\circ}$ that will cover an area of 
$100 \times 100$~km$^2$ \cite{panasyuk}, 10~m Fresnel mirrors optics  
and 5 mrad pixels \cite{khrenov}.  

This class of experiments is based on a well understood technique 
so they can be expected to be viable but it relies on significant 
technological developments. 
The technique has not yet been proven and in 
that respect the TUS (Track Ultraviolet from Space) 
project is a preparatory smaller Russian project using 1.5~m mirrors. 
It is expected to be hosted by the Resource-DK satellite to be launched 
in 2002-2003 \cite{panasyuk}. 
Upgrades of these detectors were also discussed such as MULTIOWL and 
SUPEROWL as part of a Grand Observatory in Space \cite{takahashi}. 
They intend to enhance the performance of EUSO or KLYPVE by tilting 
the system with respect to nadir, having a higher altitude orbit and 
using more mirrors to measure ZeV ($E > 10^{21}~$eV) cosmic rays and 
neutrinos. 

\subsection*{New Ideas for Composition} 

As was pointed out by many speakers composition is a fundamental 
clue in the search of the origin of the post GZK flux 
\cite{olinto,weiler,sigl}. 
The unavoidable need of using indirect detection techniques makes 
it hard to accurately determine the nature of the particles 
themselves because the differences 
in atmospheric showers developed by different primaries are rather 
subtle. Most of them can be related to either the shower 
development or to the muon content in the showers which can be 
explored by different techniques. 
Intresting new ideas were discussed for the study of the nature of the 
post-GZK particles. 

\paragraph*{Neutrinos:} 
All the detectors for measuring the post GZK flux in construction or 
planning are sensitive to neutrinos by searching for deeply penetrating 
showers \cite{croninzas,athar} 
and much use has been made of this technique to constrain 
high energy neutrino fluxes \cite{BlancoPRL}.
An interesting new idea for the detection of tau decays in the southern 
Auger observatory \cite{selvon}. 
Tau neutrinos can be expected in flavour mixing scenarios which are 
motivated by atmospheric neutrino measurements. 
The taus are produced in charged current tau neutrino interactions with 
the Andes mountains. For neutrinos in the energy range 
$10^{17}-5 \, 10^{18}$ eV the tau decays can give a detectable signal in 
the Auger tanks in spite of the solid angle being small. It was pointed 
out that it may not be possible to search for similar effects with 
fluorescence satellite detectors because their energy threshold would 
prevent the tau from decaying in the detector field of view \cite{selvon}.

\paragraph*{Photons:} 
Two new ideas were discussed for the identification of a photon component 
in the post GZK particle flux: 

The first one is due to a competition 
between the Landau-Pomeranchuck-Migdal (LPM) suppression 
of pair production and bremsstrahlung because of collective effects of 
the atmospheric nuclei and the interaction of photons with the 
geomagnetic field of the Earth. The LPM effect changes shower development 
in a dramatic fashion above an effective threshold. The shower develops 
much later becoming more elongated and a ground particle array would detect 
a much younger shower with a characteristic steep lateral profile.  
On the other hand the interaction of a photon with the 
Earth magnetic field can happen long before the start of shower 
development and has the effect of distributing the photon energy among 
lower energy secondary photons, electrons and positrons that add up to 
the primary photon energy. Provided the photon does not travel parallel 
to the magnetic field this interaction prevents the development of LPM 
showers and the atmospheric shower that follows is a scaled up version 
of the well established lower energy showers. There are however preferred 
directions along which the photon is unlikely to interact with the magnetic 
field of the Earth and the shower develops these strange LPM behaviour 
that have been carefully studied for the southern 
Auger observatory \cite{bertou}. 
The alignment of the magnetic fields gives ``holes'' for LPM showers. 
This possibility was discussed as a means to establish the photon 
fraction in cosmic rays for both particle arrays \cite{teshima} and for 
the fluorescence technique \cite{scarsi}.

A second idea was discussed in connection to the detection of horizontal 
air showers (HAS). Inclined showers induced by protons, nuclei or photons 
develop early in the atmosphere and only their muon component can be 
detected with particle arrays. The recent study of 
geomagnetic effects of the muon distributions at ground level 
\cite{zasmodel} allows the study of events above $60^ \circ$. The 
expected HAS rates have been shown to be very dependent on composition 
because of the relative content of muons in the showers induced by 
different primaries, particularly between photons and protons or nuclei. 
As a result the combined measurement of vertical and horizontal air shower 
rates provides an new handle on composition. These ideas have been applied 
to Haverah Park data which are inconsistent with a model in which photons 
are more than $42 \%$ of the cosmic ray flux above $10^{19}$ eV, which 
provides a severe constrain on models which have locally enhanced 
cosmic ray production by fragmentation \cite{zasbound,zas}. 
Since the number of muons in a 
photon shower is expected to be smaller by a large factor with respect to 
a proton (or nucleus) of the same total energy the results are fairly 
robust. These ideas can also be used 
for setting bounds on the heavy ion fraction, but the results are 
more dependent on the muon production of the assumed interaction model.     


\subsection*{The Intermediate Energy Range}

Some of the talks referred to experiments designed to measure lower 
energy particles. Low energy fluxes can be related to UHECR although 
the bulk of the radiation received in the low energy region 
is expected to be from a different origin from the UHECR \cite{gaisser}.  

\paragraph*{AMS:} 
The Anti Matter Spectrometer (AMS) also discussed in the 
conference \cite{menhaca}. 
It is an advanced detector to be flown in the ISS in 2003 for precision 
measurements of cosmic ray composition in the GeV energy range that has 
an important mexican involvement \cite{AMSmex}. 
A successful prototype has already been flown. 

\paragraph*{Ground Based GRB watch:}
Gamma Ray Bursts may be the source of UHECR. 
A review was presented of the unique Chacaltaya observatory 
located 5230 m above sea level at an atmospheric depth of only 
540~g~cm$^{-2}$. It included a 
recent proposal to transport an extensive air shower array combining 
scintillator detectors and air \v Cerenkov collectors to be used in 
combination with a 100~m$^2$ calorimeter \cite{oscar}. 
The arrangement could allow 
measurements of inelasticity by identifying the low energy subshowers 
of quasielastic events with \v Cerenkov collectors on the surface 
and could have a sensitivity to gamma rays of 16~GeV from Gamma Ray Bursts. 
An interesting proposal was made for a gamma ray watch in the 1-100 GeV range 
combining different 
ground experiments such as Chacaltaya, MILAGRO and the Auger Observatory. 
These experiments would be operated in a low threshold energy mode using 
correlated excess in the single particle counts \cite{oscar}. 

\paragraph*{Dark Matter Searches:} 

Searches for supersymmetric dark matter were reviewed by 
John Ellis \cite{ellis}. These particles can be detected via 
their annihilation in the galactic halo producing antiparticles 
which could be detected by satellites such as AMS and other 
satellites searching for gamma rays. Alternatively this possibility 
can be explored by searching for excess neutrinos in the directions 
of the Sun or the Earth center 
produced by annihilations of the supersymmetric particles 
trapped in the corresponding gravitational potentials. This method is 
the most sensitive and many specific models can be ruled out with 
10~km$^2$~y of neutrino data.  
Lastly they 
can be searched for directly by looking for their elastic scattering 
in a detector and there is a recent unconfirmed claim of evidence 
for such scattering \cite{DAMA}, but the final answer may come after LHC 
starts running in 2006. 

\section*{Epilogue: The Future of the Field}

\begin{figure}[b!] 
\centerline{\epsfig{file=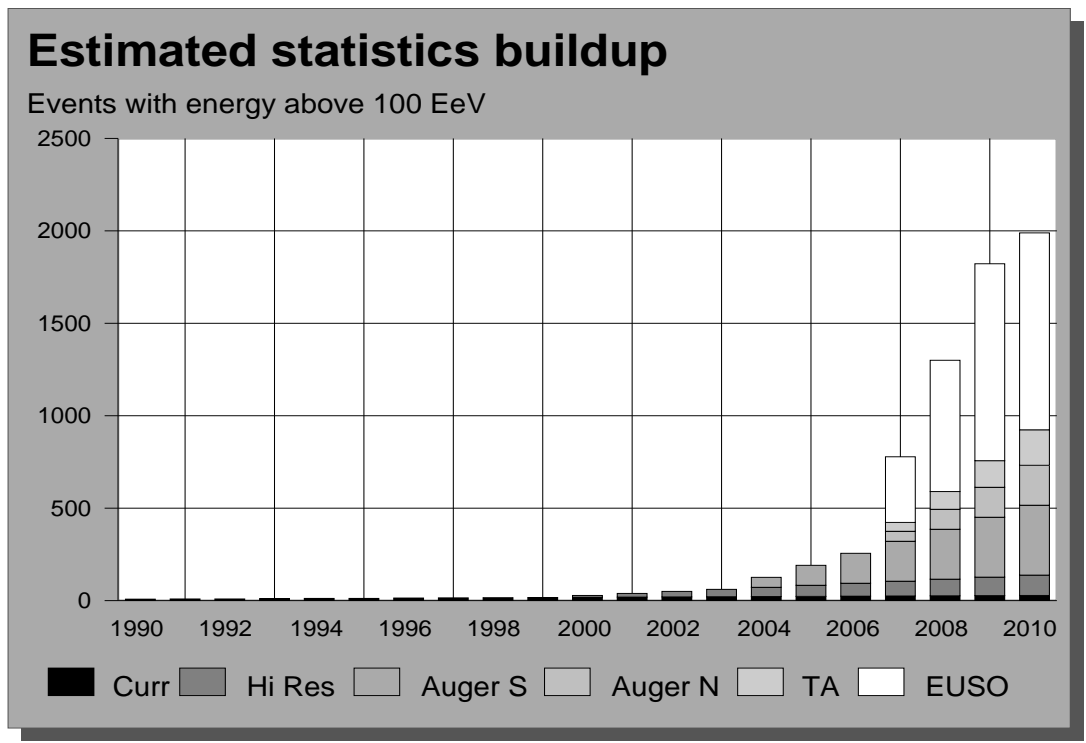,height=3.9in,width=5.7in}}
\vspace{10pt}
\caption{Accumulated events of energy exceeding $10^{20}$ eV expected to 
be detected by possible future and current experiments as a function of time: 
Current experimental data from Fig.~1 (Curr), HiRes, 
South and North Auger observatories (Auger S,N), Telescope 
Array (TA) and EUSO.   
The plotted HiRes detection rate was taken from this conference [63], 
all the other detectors are normalized to the flux detected up to now 
by AGASA. 
Data for Auger observatories assumes 9000 km$^2$~sr acceptance and 
includes all zenith angle showers, EUSO expectations are obtained 
for a 400 km orbit, considering zenith angles from 0 to 70$^ \circ$, 
assuming the project operates for three years with a $10\%$ duty cycle.}
\label{datarate}
\end{figure}

The frustratingly inconclusive statistics reached at present 
will cease to be a problem in the near future as it is 
illustrated in Fig.~\ref{datarate} where the number of events 
detected above $10^{20}~$eV is plotted as a function of time.  
The statistics will be more than doubled in two years with 
forthcoming HiRes data. 
By the end of the year 2005 we should have two years of data from the 
Auger detector and more than ten times the statistics we now have. 
The rising statistics trend will be 
continued with even more ambitious projects such as EUSO, 
KLIPVE and/or the Telescope Array. 

The new experiments will not only bring more accurate 
measurements of the post GZK spectrum, they will also bring about 
new handles on composition. All the experiments in planning 
or construction are also sensitive 
to high energy neutrinos by looking for horizontal air showers.  
In the EeV energy range the neutrino flux is expected to  
be related to the production mechanisms of the cosmic rays themselves.
Such flux will be further explored by dedicated under-water or under-ice 
neutrino experiments in construction \cite{amanda} or development \cite{antares}, quite possibly with the IceCube proposal to extend  
AMANDA to a km$^3$ size detector and also by 
searching for horizontal air showers with all the projects that are  
sensitive to ZeV Air Showers. 

The coming decade of the 2000's will bring the eagerly expected data 
that will answer many of the open questions of the post GZK flux. 
All these experiments must be regarded as a whole effort towards the  understanding of the highest energy particles in the Universe. Different 
techniques for the detection of Ultra High Energy particles will 
complement each other, resolve uncertainties which are still present and 
will open the way to many different and interesting aspects of high 
energy particle physics, astrophysics and cosmology. 

\vskip 0.5 cm
{\bf Acknowledgements:} I thank the organizers of this very 
fruitful conference in a fabulous setting under the spell of an 
impressive volcano. 
I also want to thank J.J.~Blanco-Pillado, G.~Parente, 
R.A.~V\'azquez and A.A.~Watson for many comments and suggestions 
after reading this manuscript. 
This work was supported in part by CICYT (AEN99-0589-C02-02) and 
by Xunta de Galicia (PGIDT00PXI20615PR).

\end{document}